\documentclass[aps,preprint,onecolumn,superscriptaddress]{revtex4-1}
\usepackage{amsfonts,amsmath,amssymb}
\usepackage[mathscr]{euscript}

\begin{document}
\title{Partition function of $\mathcal{N}=2$ supersymmetric gauge theory
and two-dimensional Yang-Mills theory}

\author{Xinyu Zhang}
\email{zhangxinyuphysics@gmail.com}
\affiliation{C.N. Yang Institute for Theoretical Physics, Stony Brook University,\\
Stony Brook, New York 11794-3840, USA}

\begin{abstract}
We study four-dimensional $\mathcal{N}=2$ supersymmetric $U(N)$
gauge theory with $2N$ fundamental hypermultiplets in the self-dual
$\Omega$-background. The partition function simplifies at special
points of the parameter space and is related to the partition function
of two-dimensional Yang-Mills theory on $S^{2}$. We also consider
the insertion of a Wilson loop operator in two-dimensional Yang-Mills
theory and find the corresponding operator in the four-dimensional
$\mathcal{N}=2$ gauge theory.
\end{abstract}

\preprint{YITP-SB-16-39}

\maketitle

\section{Introduction}

$\mathcal{N}=2$ supersymmetry in four dimensions imposes powerful
constraints on the low energy behavior of supersymmetric theories.
All terms with at most two derivatives and four fermions in the Wilsonian
effective action are expressed in terms of a single holomorphic quantity,
the prepotential $\mathcal{F}$, whose quantum corrections are one-loop
exact in the perturbation theory, and generated nonperturbatively
only by instantons. The exact form of the prepotential $\mathcal{F}$
was first determined for certain theories by Seiberg and Witten indirectly
based on several assumptions on the strong coupling behavior of the
theory \cite{Seiberg:1994rs,Seiberg:1994aj}. It was then extended
to more general $\mathcal{N}=2$ theories (see \cite{Tachikawa:2013kta}
for a recent review). 

It is useful to deform the supersymmetric theories by putting them
on nontrivial supergravity backgrounds \cite{Nekrasov:2002qd,Festuccia:2011ws}.
The prototypical example is the so-called $\Omega$-background \cite{Nekrasov:2002qd},
in which the theory is deformed by two parameters $\epsilon_{1},\epsilon_{2}$
parametrizing an $SO(4)$ rotation of $\mathbb{R}^{4}$. The $\Omega$-deformation
provides an IR regularization that preserves a part of the deformed
supersymmetry. The calculation of the supersymmetric partition function
is dramatically simplified and can be performed using equivariant
localization techniques. The dependence of the partition function
on the parameters $\epsilon_{1},\epsilon_{2}$ contains profound physical
information. In particular, it gives the prepotential of the low energy
effective action of the undeformed theory on $\mathbb{R}^{4}$, as
well as the couplings of the theory to the $\mathcal{N}=2$ supergravity
multiplet.

Soon after the exact computation of the partition function in the
$\Omega$-background was done, an interesting relation between supersymmetric
gauge theory and topological string theory was discovered \cite{Losev:2003py,Marshakov:2006ii}.
On the gauge theory side, we have the four-dimensional $\mathcal{N}=2$
supersymmetric $U(N)$ gauge theory with $2N-2$ fundamental hypermultiplets.
Its partition function in the self-dual $\Omega$-background simplifies
dramatically at a special point of the parameter space and is identified
with the disconnected partition function of A-type topological string
theory on $S^{2}$. The higher Casimir operators in the four-dimensional
gauge theory map to gravitational descendants of the Kähler form in
the topological string theory. It was later further generalized in
\cite{Nekrasov:2009zza} by adding $g$ adjoint hypermultiplets in
the four-dimensional gauge theory and replacing $S^{2}$ with a genus
$g$ Riemann surface. 

Inspired by the previous results, we explore the possible simplification
of the partition function of the four-dimensional $\mathcal{N}=2$
supersymmetric $U(N)$ gauge theory with $2N$ fundamental hypermultiplets
in this paper. We find that the partition function in the self-dual
$\Omega$-background at a special point of the parameter space can
be related to the partition function of two-dimensional Yang-Mills
theory on $S^{2}$ \cite{Migdal:1975zg,Cordes:1994fc}. The rank of
the gauge group of the two-dimensional theory has nothing to do with
the four-dimensional gauge group $U(N)$. 

Once the correspondence is established, one may study either side
using the information of the other side. In this paper, we consider
the Wilson loop operator in the two-dimensional Yang-Mills theory.
The exact expectation value of the Wilson loop operator has been known
for a long time. We show that inserting a Wilson loop operator in
the fundamental representation corresponds to adding a nontrivial
operator in the four-dimensional $\mathcal{N}=2$ gauge theory. The
generalization to other representations is more involved and will
be discussed in the future.

The structure of this paper is as follows. In Sec. \ref{sec:Instanton},
we review the partition function of four-dimensional $\mathcal{N}=2$
supersymmetric $U(N)$ gauge theory with $2N$ fundamental hypermultiplets
in the $\Omega$-background, and describe the $\mathscr{Y}$-observable
that will turn out to be useful in our discussion. We show that the
partition function simplifies at special points of the parameter space.
In Sec. \ref{sec:2dYM}, we show that the simplified partition function
can be related to the partition function of two-dimensional Yang-Mills
theory on $S^{2}$. We then study the effect of inserting a Wilson
loop operator in the two-dimensional Yang-Mills theory. Finally, in
Sec. \ref{sec:Dis}, we provide some further discussion.

\section{Instanton partition function of four-dimensional $\mathcal{N}=2$
gauge theory \label{sec:Instanton}}

In this paper, we are interested in the $\mathcal{N}=2$ supersymmetric
$U(N)$ gauge theory with $2N$ fundamental hypermultiplets. The Lagrangian
and the vacua are parametrized by the coupling constant $q=\exp\left(2\pi\mathrm{i}\tau\right)$,
the vacuum expectation value $\mathbf{a}=\mathrm{diag}\left(a_{1},\cdots,a_{N}\right)$
of the scalar field in the vector multiplet, and the complex masses
$\mathbf{m}=\mathrm{diag}\left(m_{1},\cdots,m_{2N}\right)$ of the
matter hypermultiplets. We refer to \cite{Nekrasov:2012xe} for a
detailed analysis and references for the supersymmetric partition
function of very general $\mathcal{N}=2$ supersymmetric gauge theories
in the $\Omega$-background.

\subsection{Partition function in the self-dual $\Omega$-background}

Let us first recall the partition function of the four-dimensional
$\mathcal{N}=2$ gauge theory in the $\Omega$-background \cite{Nekrasov:2002qd}.
The $\Omega$-background breaks the translational invariance by deforming
the theory in a rotationally covariant way, with parameters $\epsilon_{1},\epsilon_{2}$.
In the following, we always set $\epsilon_{1}=-\hbar$, $\epsilon_{2}=\hbar$. 

The supersymmetric partition function of $\mathcal{N}=2$ theory consists
of three parts: the classical, the one-loop, and the instanton parts,

\begin{equation}
Z\left(\mathbf{a},\mathbf{m},q;\hbar\right)=Z^{\mathrm{classical}}\left(\mathbf{a},q;\hbar\right)Z^{1-\mathrm{loop}}\left(\mathbf{a},\mathbf{m};\hbar\right)Z^{\mathrm{instanton}}\left(\mathbf{a},\mathbf{m},q;\hbar\right).
\end{equation}
The classical part is simply 
\begin{equation}
Z^{\mathrm{classical}}\left(\mathbf{a},q;\hbar\right)=q^{\frac{1}{2\hbar^{2}}\sum_{\alpha=1}^{N}a_{\alpha}^{2}}.
\end{equation}
The one-loop part is given as a product of contributions from the
vector multiplet and the matter hypermultiplets using the Barnes double
gamma function. The one-loop contribution of a vector multiplet is
\begin{equation}
Z_{\mathrm{vector}}^{1-\mathrm{loop}}\left(\mathbf{a};\hbar\right)=\prod_{1\leq i<j\leq N}\left[\Gamma_{2}\left(a_{i}-a_{j}+\hbar|\hbar,-\hbar\right)\Gamma_{2}\left(a_{i}-a_{j}-\hbar|\hbar,-\hbar\right)\right]^{-1},
\end{equation}
while the one-loop contribution of fundamental hypermultiplets is
\begin{equation}
Z_{\mathrm{fund}}^{1-\mathrm{loop}}\left(\mathbf{a},m;\hbar\right)=\prod_{i=1}^{N}\prod_{f=1}^{2N}\Gamma_{2}\left(a_{i}-m_{f}|\hbar,-\hbar\right).
\end{equation}
The instanton partition function is defined as an equivariant integral
over the instanton moduli space. Applying the equivariant localization
method, the integral can be reduced to a sum over contributions of
the fixed points of the moduli space. There is a one-to-one correspondence
between the fixed points and colored partitions $\Lambda=\left(\lambda^{(\alpha)}\right)_{\alpha=1}^{N}$,
with each partition $\lambda^{(\alpha)}$ being a weakly decreasing
sequence of non-negative integers,
\begin{equation}
\lambda^{(\alpha)}=\left(\lambda_{1}^{(\alpha)}\geq\lambda_{2}^{(\alpha)}\geq\cdots\geq\lambda_{\ell\left(\lambda^{(\alpha)}\right)}^{(\alpha)}>\lambda_{\ell\left(\lambda^{(\alpha)}\right)+1}^{(\alpha)}=\cdots=0\right),\label{eq:partition}
\end{equation}
whose size is denoted to be $|\lambda^{(\alpha)}|=\sum_{i}\lambda_{i}^{(\alpha)}$.
Accordingly the instanton partition function becomes a statistical
model of random partitions \cite{Nekrasov:2002qd},
\begin{equation}
Z^{\mathrm{instanton}}\left(\mathbf{a},\mathbf{m},q;\hbar\right)=\sum_{\Lambda}q^{|\Lambda|}\mu_{\Lambda}\left(\mathbf{a},\mathbf{m};\hbar\right),\label{eq:Z}
\end{equation}
where $|\Lambda|=\sum_{\alpha=1}^{N}|\lambda^{(\alpha)}|$. The contribution
to the measure of a vector multiplet is given by
\begin{equation}
\mu_{\Lambda\mathrm{vector}}\left(\mathbf{a};\hbar\right)=\prod_{\left(\alpha,i\right)\neq\left(\beta,j\right)}\frac{a_{\alpha}-a_{\beta}+\hbar\left(\lambda_{i}^{(\alpha)}-\lambda_{j}^{(\beta)}+j-i\right)}{a_{\alpha}-a_{\beta}+\hbar\left(j-i\right)},\label{eq:inst-v}
\end{equation}
and the contribution to the measure of fundamental hypermultiplets
is 
\begin{eqnarray}
\mu_{\Lambda\mathrm{fund}}\left(\mathbf{a},\mathbf{m};\hbar\right) & = & \prod_{\alpha=1}^{N}\prod_{f=1}^{2N}\prod_{\square\in\lambda^{(\alpha)}}\left(c_{\square}-m_{f}\right)\nonumber \\
 & = & \hbar^{2N|\Lambda|}\prod_{\alpha=1}^{N}\prod_{f=1}^{2N}\prod_{i}\frac{\Gamma\left(\frac{a_{\alpha}-m_{f}}{\hbar}+1+\lambda_{i}^{(\alpha)}-i\right)}{\Gamma\left(\frac{a_{\alpha}-m_{f}}{\hbar}+1-i\right)},\label{eq:inst-fund}
\end{eqnarray}
where for each box $\square=(i,j)\in\lambda^{(\alpha)}$, we define
its content as
\begin{equation}
c_{\square}=a_{\alpha}+\epsilon_{1}\left(i-1\right)+\epsilon_{2}\left(j-1\right).\label{eq:content}
\end{equation}
The contribution to the measure of an antifundamental hypermultiplet
with mass $m$ is equal to the contribution to the measure of a fundamental
hypermultiplet with mass $-m$ in the self-dual $\Omega$-background.

For the undeformed theory on $\mathbb{R}^{4}$, we can perturb the
theory by adding gauge-invariant chiral operators to the ultraviolet
prepotential, while keeping the ultraviolet antiprepotential unchanged,
\begin{equation}
\bar{\mathcal{F}}^{\mathrm{UV}}=\frac{\bar{\tau}}{2}\mathrm{Tr}\bar{\Phi}^{2}.
\end{equation}
For example, we can add single-trace operators, 
\begin{equation}
\mathcal{F}^{\mathrm{UV}}\to\frac{\tau}{2}\mathrm{Tr}\Phi^{2}+\sum_{j=2}^{\infty}\frac{\tau_{j}}{j}\mathrm{Tr}\Phi^{j},
\end{equation}
which get deformed in the $\Omega$-background. The localization computation
still works, and the partition function becomes
\begin{equation}
Z\left(\mathbf{a},\mathbf{m},q;\tau;\hbar\right)=Z^{\mathrm{classical}}\left(\mathbf{a},q;\hbar\right)Z^{1-\mathrm{loop}}\left(\mathbf{a},\mathbf{m};\hbar\right)\sum_{\Lambda}q^{|\Lambda|}\mu_{\Lambda}\left(\mathbf{a},\mathbf{m};\hbar\right)\exp\left(\frac{1}{\hbar^{2}}\sum_{j=2}^{\infty}\frac{\tau_{j}}{j}\mathrm{ch}_{j}\left(\mathbf{a},\Lambda\right)\right).\label{eq:Ztau}
\end{equation}
Here $\mathrm{ch}_{j}\left(\mathbf{a},\Lambda\right)=\sum_{\alpha=1}^{N}\mathrm{ch}_{j}\left(a_{\alpha},\lambda^{(\alpha)}\right)$,
with
\begin{equation}
\mathrm{ch}_{j}\left(a,\lambda\right)=a^{j}+\sum_{i=1}^{\infty}\left(\left(a+\hbar\left(\lambda_{i}+1-i\right)\right)^{j}-\left(a+\hbar\left(\lambda_{i}-i\right)\right)^{j}-\left(a+\hbar\left(1-i\right)\right)^{j}+\left(a-\hbar i\right)^{j}\right).
\end{equation}
For example, 
\begin{eqnarray}
\mathrm{ch}_{2}\left(a,\lambda\right) & = & a^{2}+2\hbar^{2}|\lambda|,\\
\mathrm{ch}_{3}\left(a,\lambda\right) & = & a^{3}+6\hbar^{2}a|\lambda|+3\hbar^{3}\sum_{i}\lambda_{i}\left(\lambda_{i}+1-2i\right).
\end{eqnarray}

Multitrace operators can also be added and can be analyzed using the
Hubbard-Stratonovich transformation. The full set of gauge-invariant
chiral operators can be expressed as 
\begin{equation}
\mathcal{F}^{\mathrm{UV}}\to\frac{\tau}{2}\mathrm{Tr}\Phi^{2}+\sum_{\vec{k}}^{\infty}t_{\vec{k}}\prod_{j=1}^{\infty}\frac{1}{k_{j}!}\left(\frac{1}{j}\mathrm{Tr}\Phi^{j}\right)^{k_{j}},\quad\vec{k}=\left(k_{1},k_{2},\cdots\right),\label{eq:perturb}
\end{equation}
and the partition function is deformed to be
\begin{eqnarray}
Z\left(\mathbf{a},\mathbf{m},q;\mathbf{t};\hbar\right) & = & Z^{\mathrm{classical}}\left(\mathbf{a},q;\hbar\right)Z^{1-\mathrm{loop}}\left(\mathbf{a},\mathbf{m};\hbar\right)\times\nonumber \\
 & \times & \text{\ensuremath{\sum_{\Lambda}q^{|\Lambda|}\mu_{\Lambda}\left(\mathbf{a},\mathbf{m};\hbar\right)\exp\left(\frac{1}{\hbar^{2}}\sum_{\vec{k}}^{\infty}t_{\vec{k}}\prod_{j=1}^{\infty}\frac{1}{k_{j}!}\left(\frac{1}{j}\mathrm{ch}_{j}\left(a,\lambda\right)\right)^{k_{j}}\right)}.}\label{eq:Zt}
\end{eqnarray}

\subsection{$\mathscr{Y}$-observable }

With the identification of the instanton partition function with a
statistical model (\ref{eq:Z}), we can compute the expectation value
of observables in the $\Omega$-background as
\begin{equation}
\langle\mathcal{O}\rangle=\frac{\sum_{\Lambda}q^{|\Lambda|}\mu_{\Lambda}\mathcal{O}[\Lambda]}{\sum_{\Lambda}q^{|\Lambda|}\mu_{\Lambda}},\label{eq:vevO}
\end{equation}
where $\mathcal{O}[\Lambda]$ is the value of $\mathcal{O}$ at the
fixed point labeled by $\Lambda$.

An important observable in the analysis of nonperturbative information
of four-dimensional $\mathcal{N}=2$ gauge theory is the $\mathscr{Y}$-observable,
which is defined using the gauge-invariant polynomials of the adjoint
scalar field $\phi$ in the vector multiplet, evaluated at the fixed
point of the rotational symmetry $SO(4)$,
\begin{equation}
\mathscr{Y}(x)=x^{N}\exp\left(-\sum_{j=1}^{\infty}\frac{1}{jx^{j}}\mathrm{Tr}\left(\phi(0)\right)^{j}\right).
\end{equation}
Classically, it is given by
\begin{equation}
\mathscr{Y}(x)^{\mathrm{classical}}=\det\left(x-\phi(0)\right)=\prod_{\alpha=1}^{N}\left(x-a_{\alpha}\right).
\end{equation}
However, there are quantum corrections due to instantons. Denote the
outer and the inner boundaries of the partition $\lambda$ as $\partial_{+}\lambda$
and $\partial_{-}\lambda$, respectively. The value of $\mathscr{Y}(x)$
in the self-dual $\Omega$-background at the fixed point labeled by
$\Lambda$ is \cite{Nekrasov:2015wsu}
\begin{eqnarray}
\mathscr{Y}(x)[\Lambda] & = & \prod_{\alpha=1}^{N}\frac{\prod_{\boxplus\in\partial_{+}\lambda^{(\alpha)}}\left(x-c_{\boxplus}\right)}{\prod_{\boxminus\in\partial_{-}\lambda^{(\alpha)}}\left(x-c_{\boxminus}\right)}\nonumber \\
 & = & \prod_{\alpha=1}^{N}\prod_{i}^{\infty}\frac{x-a_{\alpha}-\hbar\left(\lambda_{i}^{(\alpha)}-i+1\right)}{x-a_{\alpha}-\hbar\left(\lambda_{i}^{(\alpha)}-i\right)}.\label{eq:Y}
\end{eqnarray}
Notice that the expression (\ref{eq:Y}) is highly redundant, and
there can be many cancellations between the numerator and the denominator.
For example, the contribution from the box $\left(n+1,\lambda_{n+1}^{(\alpha)}+1\right)\in\widetilde{\partial_{+}\lambda^{(\alpha)}}$
cancels the contribution from the box $\left(n,\lambda_{n}^{(\alpha)}\right)\in\widetilde{\partial_{-}\lambda^{(\alpha)}}$
for $n>\ell(\lambda^{(\alpha)})$. Hence, $\mathscr{Y}(x)[\Lambda]$
does not change if we truncate the range of the index $i$ to $1\leq i\leq n$
for an arbitrary integer $n\geq\ell(\lambda^{(\alpha)})$. 

\subsection{Simplification of partition function}

Up to this point we assumed that the expectation values $a_{1},\cdots,a_{N}$
and masses $m_{1},\cdots,m_{2N}$ are generic. Then the partition
function (\ref{eq:Z}) contains an infinite sum over colored partitions.
For a special value of the masses, the partitions $\Lambda$ that
we sum over can be constrained. As a result, the partition function
(\ref{eq:Z}) gets simplified.

It is easy to see that if $a_{\alpha}=m_{f}$ for some $\alpha\in\left\{ 1,2,\cdots N\right\} $
and $f\in\left\{ 1,2,\cdots,2N\right\} $, then $\lambda^{(\alpha)}=\emptyset$;
otherwise (\ref{eq:inst-fund}) is zero. Therefore, if we choose a
particular point on the parameter space
\begin{equation}
a_{\alpha}=m_{2\alpha-1}=m_{2\alpha},\quad\alpha=1,\cdots,N,\label{eq:trivial}
\end{equation}
the partitions $\lambda^{(\alpha)}=\emptyset$ for all $\alpha=1,2,\cdots,N$,
and the instanton partition function is trivially $1$. This simplification
of the instanton partition function has been known for a long time.
Physically, when one of the $a_{\alpha}$'s is equal to two masses,
two of the hypermultiplets become massless, and can be Higgsed so
that the $U(N)$ theory with $2N$ flavors is reduced to a $U(N-1)$
theory with $2N-2$ flavors. However, the instanton partition function
will not change since it is a Coulomb-branch quantity which is independent
of the manipulation on the hypermultiplet side.

Now let us relax the condition (\ref{eq:trivial}) a little bit. We
still fix
\begin{equation}
a_{\alpha}=m_{2\alpha-1}=m_{2\alpha},\quad\alpha=2,\cdots,N,\label{eq:condition2}
\end{equation}
so that the partitions $\lambda^{(\alpha)}=\emptyset$ for $\alpha=2,\cdots,N$.
We effectively reduce the $U(N)$ gauge theory with $2N$ fundamental
hypermultiplets to the $U(1)$ theory with two fundamental hypermultiplets.
At the same time, we choose
\begin{equation}
a_{1}=m_{1}+n\hbar=m_{2}+n\hbar,\label{eq:condition1}
\end{equation}
where $n$ is a positive integer. We see from (\ref{eq:inst-fund})
that if $\lambda_{n+1}^{(1)}\geq1$, then the contribution of the
box $\square=(n+1,1)\in\lambda^{(1)}$ makes $\mu_{\Lambda\mathrm{fund}}$
vanish. Hence, the length of the partition $\lambda^{(1)}$ is at
most $n$. We can set the length of the partition $\lambda^{(1)}$
to be $n$ by adding zeros to the end of the partition if its precise
length is less than $n$. In this case, the measure in the instanton
partition function simplifies. 

The case $n=1$ is special, since now $\lambda^{(1)}$ is no longer
a two-dimensional partition. The measure of the vector multiplet completely
cancels the measure of the fundamental hypermultiplets, and the instanton
partition function is
\begin{equation}
Z^{\mathrm{instanton}}=\sum_{\lambda_{1}^{(1)}=0}^{\infty}q^{\lambda_{1}^{(1)}}=\frac{1}{1-q}.
\end{equation}
In the following, we always assume that $n\geq2$. In this case, the
measure of the vector multiplet (\ref{eq:inst-v}) becomes
\begin{eqnarray}
\mu_{\Lambda\mathrm{vector}} & = & \left(\prod_{i\neq j}\frac{\hbar\left(\lambda_{i}^{(1)}-\lambda_{j}^{(1)}+j-i\right)}{\hbar\left(j-i\right)}\right)\left(\prod_{\beta=2}^{N}\prod_{i,j}\frac{a_{1}-a_{\beta}+\hbar\left(\lambda_{i}^{(1)}+j-i\right)}{a_{1}-a_{\beta}+\hbar\left(j-i\right)}\right)^{2}\nonumber \\
 & = & \left(\prod_{1\leq i<j\leq n}\frac{\lambda_{i}^{(1)}-\lambda_{j}^{(1)}+j-i}{j-i}\right)^{2}\left(\prod_{i=1}^{n}\frac{\Gamma\left(n+1-i\right)}{\hbar^{\lambda_{i}^{(1)}}\Gamma\left(n+1+\lambda_{i}^{(1)}-i\right)}\right)^{2}\times\nonumber \\
 & \times & \left(\prod_{\beta=2}^{N}\prod_{i=1}^{n}\frac{\Gamma\left(\frac{a_{1}-a_{\beta}}{\hbar}-i+1\right)}{\hbar^{\lambda_{i}^{(1)}}\Gamma\left(\frac{a_{1}-a_{\beta}}{\hbar}-i+\lambda_{i}^{(1)}+1\right)}\right)^{2},
\end{eqnarray}
while the measure of the fundamental hypermultiplets (\ref{eq:inst-fund})
becomes
\begin{eqnarray}
\mu_{\Lambda\mathrm{fund}} & = & \prod_{f=1}^{2N}\prod_{i=1}^{n}\frac{\Gamma\left(\frac{a_{1}-m_{f}}{\hbar}+1+\lambda_{i}^{(1)}-i\right)}{\Gamma\left(\frac{a_{1}-m_{f}}{\hbar}+1-i\right)}\nonumber \\
 & = & \hbar^{2N|\lambda^{(1)}|}\left(\prod_{i=1}^{n}\frac{\Gamma\left(n+1+\lambda_{i}^{(1)}-i\right)}{\Gamma\left(n+1-i\right)}\right)^{2}\prod_{\alpha=2}^{N}\left(\prod_{i=1}^{n}\frac{\Gamma\left(\frac{a_{1}-a_{\alpha}}{\hbar}+1+\lambda_{i}^{(1)}-i\right)}{\Gamma\left(\frac{a_{1}-a_{\alpha}}{\hbar}+1-i\right)}\right)^{2}.
\end{eqnarray}
After many cancellations between $\mu_{\Lambda\mathrm{vector}}$ and
$\mu_{\Lambda\mathrm{fund}}$, the remaining measure is
\begin{equation}
\mu_{\Lambda}=\mu_{\Lambda\mathrm{vector}}\mu_{\Lambda\mathrm{fund}}=\left(\prod_{1\leq i<j\leq n}\frac{\lambda_{i}^{(1)}-\lambda_{j}^{(1)}+j-i}{j-i}\right)^{2}.\label{eq:mu}
\end{equation}

In this case, the $\mathscr{Y}$-observable (\ref{eq:Y}) also simplifies,
\begin{eqnarray}
\mathscr{Y}(x)[\Lambda] & = & \frac{\prod_{i=1}^{n+1}\left(x-a_{1}-\hbar\left(\lambda_{i}^{(1)}+1-i\right)\right)}{\prod_{i=1}^{n}\left(x-a_{1}-\hbar\left(\lambda_{i}^{(1)}-i\right)\right)}\nonumber \\
 & = & \left(x-a_{1}+n\hbar\right)\prod_{i=1}^{n}\frac{\left(x-a_{1}-\hbar\left(\lambda_{i}^{(1)}+1-i\right)\right)}{\left(x-a_{1}-\hbar\left(\lambda_{i}^{(1)}-i\right)\right)}.
\end{eqnarray}

As we see, at the point (\ref{eq:condition2}) (\ref{eq:condition1})
of the parameter space, the instanton partition function is independent
of the gauge group rank $N$, and the difference for different $N$
values in the full partition function is an overall constant which
is irrelevant to our discussion. Therefore, we shall concentrate on
the case $N=1$ in the following discussion and drop some of the subscripts
$1$. Notice that the $U(1)$ gauge theory with two fundamental hypermultiplets
is nontrivial due to the inexplicit noncommutative deformation.

\section{Relation to two-dimensional Yang-Mills theory \label{sec:2dYM}}

In this section, we shall relate the partition function discussed
in Sec. \ref{sec:Instanton} to the partition function of two-dimensional
Yang-Mills theory on $S^{2}$. 

\subsection{Partition function of two-dimensional Yang-Mills theory}

Two-dimensional Yang-Mills theory is an exactly solvable model and
has been extensively studied from many different points of view (see
\cite{Cordes:1994fc} for a review). Its partition function on a Riemann
surface $\Sigma$ of genus $g$ is defined as
\begin{equation}
Z_{\Sigma}^{\mathrm{YM2}}\left(\varepsilon,\mathcal{A}(\Sigma),G\right)=\frac{1}{\mathrm{Vol}(G)}\int\mathcal{D}A\mathcal{D}\phi\exp\left(\mathrm{i}\int_{\Sigma}\mathrm{Tr}\phi F_{A}+\frac{\varepsilon}{2}\int_{\Sigma}d\mu\mathrm{Tr}\phi^{2}\right),\label{eq:YM2}
\end{equation}
where $\varepsilon$ is the coupling constant, $\mathcal{A}(\Sigma)$
is the area of the Riemann surface $\Sigma$, and $\mathrm{Tr}$ denotes
the invariant, negative-definite quadratic form on the Lie algebra
$\mathfrak{g}$ of the gauge group $G$. The partition function (\ref{eq:YM2})
can be expressed as a sum over all finite-dimensional irreducible
representations $R$ of the gauge group $G$ \cite{Migdal:1975zg,Rusakov:1990rs,Witten:1991we},
\begin{equation}
Z_{\Sigma}^{\mathrm{YM2}}\left(\varrho,G\right)=e^{-\beta\left(2-2g\right)-\gamma\varepsilon\mathcal{A}(\Sigma)}\sum_{R}\left(\dim R\right)^{2-2g}\exp\left(-\frac{\varrho}{2}C_{2}(R)\right),\label{eq:Migdal_formula}
\end{equation}
where the prefactor is the regularization-dependent ambiguity, $\dim R$
is the dimension of the representation $R$, $C_{2}(R)$ is the quadratic
Casimir of the representation $R$, and $\varrho=\varepsilon\mathcal{A}(\Sigma)$
is the dimensionless coupling constant.

\subsection{Matching the parameters }

We would like to find the precise relation between the partition function
(\ref{eq:Zt}) and the partition function of two-dimensional Yang-Mills
theory (\ref{eq:Migdal_formula}), both for the group $SU(n)$ and
for the group $U(n)$.

\subsubsection{$SU(n)$ theory}

For the group $G=SU(n)$, the irreducible representations $R$ are
parametrized by the partition $\left(\lambda_{1}\geq\lambda_{2}\geq\cdots\geq\lambda_{n-1}\geq\lambda_{n}=0\right)$.
The dimension and the quadratic Casimir of the representation $R$
are
\begin{align}
\dim R & =\prod_{1\leq i<j\leq n}\frac{\lambda_{i}-\lambda_{j}+j-i}{j-i},\label{eq:dimR}\\
C_{2}(R) & =\sum_{i=1}^{n}\lambda_{i}\left(\lambda_{i}-2i+1\right)+n|\lambda|-\frac{|\lambda|^{2}}{n}.\label{eq:C2R}
\end{align}
We see that both the dimension and the quadratic Casimir are independent
of the overall shift of $\lambda$'s. Therefore, the difference between
the summation over $\lambda_{1}\geq\lambda_{2}\geq\cdots\geq\lambda_{n-1}\geq\lambda_{n}\geq0$
and $\lambda_{1}\geq\lambda_{2}\geq\cdots\geq\lambda_{n-1}\geq\lambda_{n}=0$
in the partition function is merely an irrelevant overall constant.

To identify the partition function of two-dimensional $SU(n)$ Yang-Mills
theory on $S^{2}$ with the partition function of the four-dimensional
$\mathcal{N}=2$ $U(1)$ gauge theory with two fundamental hypermultiplets
at the degenerate point of the parameter space, we need to set $a=0$
and turn on operators with couplings $t_{0,1}$, $t_{0,2}$ and $t_{0,0,1}$
in (\ref{eq:Zt}). The partition function becomes
\begin{eqnarray}
 &  & Z\left(a=0,m_{1}=m_{2}=-n\hbar,q;\tau;\hbar\right)\nonumber \\
 & = & \Gamma_{2}\left(n\hbar|\hbar,-\hbar\right)^{2}\sum_{\lambda}q^{|\lambda|}\left(\prod_{1\leq i<j\leq n}\frac{\lambda_{i}-\lambda_{j}+j-i}{j-i}\right)^{2}\times\\
 &  & \times\exp\left\{ \frac{1}{\hbar^{2}}\left[\frac{t_{0,1}}{2}\mathrm{ch}_{2}\left(0,\lambda\right)+\frac{t_{0,2}}{8}\left(\mathrm{ch}_{2}\left(0,\lambda\right)\right)^{2}+\frac{t_{0,0,1}}{3}\mathrm{ch}_{3}\left(0,\lambda\right)\right]\right\} \nonumber \\
 & = & \Gamma_{2}\left(n\hbar|\hbar,-\hbar\right)^{2}\sum_{\lambda}q^{|\lambda|}\left(\prod_{1\leq i<j\leq n}\frac{\lambda_{i}-\lambda_{j}+j-i}{j-i}\right)^{2}\times\nonumber \\
 &  & \times\exp\left\{ \left[t_{0,1}|\lambda|+\frac{t_{0,2}\hbar^{2}}{2}|\lambda|^{2}+t_{0,0,1}\hbar\sum_{i}\lambda_{i}\left(\lambda_{i}+1-2i\right)\right]\right\} .
\end{eqnarray}
Ignoring the unimportant prefactor coming from the one-loop contribution,
the partition function is equal to the partition function of two-dimensional
Yang-Mills theory on $S^{2}$ (\ref{eq:Migdal_formula}) with gauge
group $SU(n)$ when
\begin{eqnarray}
 &  & \log(q)|\lambda|+t_{0,1}|\lambda|+\frac{t_{0,2}\hbar^{2}}{2}|\lambda|^{2}+t_{0,0,1}\hbar\sum_{i}\lambda_{i}\left(\lambda_{i}+1-2i\right)\nonumber \\
 & = & -\frac{\varrho}{2}\left(\sum_{i=1}^{n}\lambda_{i}\left(\lambda_{i}-2i+1\right)+n|\lambda|-\frac{|\lambda|^{2}}{n}\right),
\end{eqnarray}
which gives
\begin{equation}
t_{0,1}=-\frac{\varrho n}{2}-\log\left(q\right),\quad t_{0,2}=\frac{\varrho}{n\hbar^{2}},\quad t_{0,0,1}=-\frac{\varrho}{2\hbar}.
\end{equation}

\subsubsection{$U(n)$ theory}

For the group $U(n)$, the irreducible representations $\mathcal{R}$
are parametrized by $n$ integers $\left(\mu_{1}\geq\mu_{2}\geq\cdots\geq\mu_{n-1}\geq\mu_{n}\right)$
without positivity restriction. It is convenient to use the decomposition
of the representation $\mathcal{R}$ of $U(n)$ in terms of representation
$R$ of $SU(n)$ and the $U(1)$ charge $p$,
\begin{eqnarray}
\mu_{i} & = & \lambda_{i}+r,\quad i=1,2,\cdots,n-1\nonumber \\
\mu_{n} & = & r,\nonumber \\
p & = & |\lambda|+nr,\quad r\in\mathbb{Z}.
\end{eqnarray}
The dimension of representation $\mathcal{R}$ of group $U(n)$ has
the same form (\ref{eq:dimR}) as the group $SU(n)$, while the quadratic
Casimir is given by
\begin{equation}
C_{2}\left(\mathcal{R}\right)=C_{2}\left(R\right)+\frac{p^{2}}{n}=\sum_{i=1}^{n}\lambda_{i}\left(\lambda_{i}-2i+1\right)+\left(n+2r\right)|\lambda|+nr^{2}.
\end{equation}

To relate the four-dimensional theory to two-dimensional Yang-Mills
theory with gauge group $U(n)$, we no longer need to turn on the
double-trace operators. Instead, we turn on operators with parameter
$\tau_{2}$ and $\tau_{3}$ in (\ref{eq:Ztau}),
\begin{eqnarray}
 &  & Z\left(a,\mathbf{m},q;\tau;\hbar\right)\nonumber \\
 & = & \Gamma_{2}\left(n\hbar|\hbar,-\hbar\right)^{2}\sum_{\lambda}\left(\prod_{1\leq i<j\leq n}\frac{\lambda_{i}-\lambda_{j}+j-i}{j-i}\right)^{2}\times\nonumber \\
 &  & \times\exp\left[\left(\tau_{2}+\log(q)\right)\left(\frac{a^{2}}{2\hbar^{2}}+|\lambda|\right)+\tau_{3}\left(\frac{a^{3}}{3\hbar^{2}}+2a|\lambda|+\hbar\sum_{i}\lambda_{i}\left(\lambda_{i}+1-2i\right)\right)\right].
\end{eqnarray}
We now set 
\begin{equation}
a=m_{1}+n\hbar=m_{2}+n\hbar=r\hbar,
\end{equation}
where $r\in\mathbb{Z}$. Ignoring the irrelevant prefactor coming
from the one-loop contribution, the partition function becomes
\begin{eqnarray}
 &  & Z\left(r\hbar,(r-n)\hbar,q;\tau;\hbar\right)\nonumber \\
 & = & \sum_{\lambda}\left(\prod_{1\leq i<j\leq n}\frac{\lambda_{i}-\lambda_{j}+j-i}{j-i}\right)^{2}\times\nonumber \\
 &  & \times\exp\left[\left(\tau_{2}+\log(q)\right)\left(\frac{r^{2}}{2}+|\lambda|\right)+\tau_{3}\hbar\left(\frac{r^{3}}{3}+2r|\lambda|+\sum_{i}\lambda_{i}\left(\lambda_{i}+1-2i\right)\right)\right].
\end{eqnarray}
Now we consider the sum over $r\in\mathbb{Z}$ with a possible weight
depending on $r$,
\begin{eqnarray}
 &  & \sum_{r\in\mathbb{Z}}\exp\left(-f_{2}r^{2}-f_{3}r^{3}\right)Z\left(r\hbar,(r-n)\hbar,q;\tau;\hbar\right)\nonumber \\
 & = & \sum_{r\in\mathbb{Z}}\sum_{\lambda}\left(\prod_{1\leq i<j\leq n}\frac{\lambda_{i}-\lambda_{j}+j-i}{j-i}\right)^{2}\times\nonumber \\
 &  & \times\exp\left[\left(\tau_{2}+\log(q)\right)\left(\frac{r^{2}}{2}+|\lambda|\right)+\tau_{3}\hbar\left(\frac{r^{3}}{3}+2r|\lambda|+\sum_{i}\lambda_{i}\left(\lambda_{i}+1-2i\right)\right)-f_{2}r^{2}-f_{3}r^{3}\right]
\end{eqnarray}
which is equal to the partition function of two-dimensional Yang-Mills
theory on $S^{2}$ (\ref{eq:Migdal_formula}) with gauge group $U(n)$
when
\begin{eqnarray}
 &  & \tau_{3}\hbar\sum_{i}\lambda_{i}\left(\lambda_{i}+1-2i\right)+\left(\tau_{2}+\log(q)\right)|\lambda|+2\tau_{3}\hbar|\lambda|r+\left(\frac{\tau_{2}+\log(q)}{2}-f_{2}\right)r^{2}+\left(\frac{\tau_{3}\hbar}{3}-f_{3}\right)r^{3}\nonumber \\
 & = & -\frac{\varrho}{2}\left[\sum_{i=1}^{n}\lambda_{i}\left(\lambda_{i}-2i+1\right)+n|\lambda|+2r|\lambda|+nr^{2}\right],
\end{eqnarray}
which gives that
\begin{equation}
\tau_{2}=-\frac{\varrho n}{2}-\log\left(q\right),\quad\tau_{3}=-\frac{\varrho}{2\hbar},\quad f_{2}=\frac{\varrho n}{4},\quad f_{3}=-\frac{\varrho}{6\hbar}.
\end{equation}
Therefore, we have the relation
\begin{equation}
\sum_{r\in\mathbb{Z}}\exp\left(-\frac{\varrho n}{4}r^{2}+\frac{\varrho}{6\hbar}r^{3}\right)Z\left(r\hbar,(r-n)\hbar,q;\tau_{2}=-\frac{\varrho n}{2}-\log\left(q\right),\tau_{3}=-\frac{\varrho}{2\hbar};\hbar\right)=Z_{S^{2}}^{\mathrm{YM2}}\left(\varrho,U(n)\right).
\end{equation}

\subsection{Wilson loop operator in two-dimensional Yang-Mills theory}

The correspondence was hitherto at the level of the partition functions.
We would like to deepen it by studying the Wilson loop operator in
the two-dimensional Yang-Mills theory.

Suppose that a loop $\Gamma$ decomposes $S^{2}$ into two disjoint
connected components $\Sigma_{1}$ and $\Sigma_{2}$. Associated to
the curve $\Gamma$ we have a representation $R_{\Gamma}$ of the
gauge group and we define a Wilson loop operator
\begin{equation}
W\left(\Gamma,R_{\Gamma}\right)=\mathrm{Tr}_{R_{\Gamma}}P\exp\oint_{\Gamma}A.
\end{equation}
The expectation value of the Wilson loop operator $W\left(\Gamma,R_{\Gamma}\right)$
is given by
\begin{eqnarray}
\langle W\left(\Gamma,R_{\Gamma}\right)\rangle^{\mathrm{YM2}} & = & Z_{S^{2}}^{\mathrm{YM2}}\left(\varepsilon\mathcal{A}\left(\Sigma_{1}\cup\Sigma_{2}\right)\right)^{-1}\sum_{R_{1},R_{2}}\left(\dim R_{1}\right)\left(\dim R_{2}\right)\times\nonumber \\
 & \times & \exp\left(-\frac{\varepsilon\mathcal{A}(\Sigma_{1})}{2}C_{2}(R_{1})-\frac{\varepsilon\mathcal{A}(\Sigma_{2})}{2}C_{2}(R_{2})\right)\mathfrak{N}\left(R_{1}\otimes R_{\Gamma},R_{2}\right)
\end{eqnarray}
where $\mathfrak{N}\left(R_{1}\otimes R_{\Gamma},R_{2}\right)$ is
the fusion number defined by the decomposition of a tensor product
into irreducible representations:
\begin{equation}
R_{1}\otimes R_{\Gamma}=\bigoplus_{R_{2}}\mathfrak{N}\left(R_{1}\otimes R_{\Gamma},R_{2}\right)R_{2}.
\end{equation}
In this paper, we are interested in the simple case that $R_{\Gamma}$
is the fundamental representation. The fusion number is $1$ if the
Young diagram associated to $R_{2}$ is obtained by adding a box in
the Young diagram associated to $R_{1}$, and $0$ otherwise. We can
make an analogy with (\ref{eq:vevO}) and write
\begin{equation}
\langle W\left(\Gamma,\square\right)\rangle^{\mathrm{YM2}}=Z_{S^{2}}^{\mathrm{YM2}}\left(\varepsilon\mathcal{A}\left(\Sigma_{1}\cup\Sigma_{2}\right)\right)^{-1}\sum_{R}\left(\dim R\right)^{2}\exp\left(-\frac{\varepsilon\mathcal{A}(\Sigma_{1}\cup\Sigma_{2})}{2}C_{2}(R)\right)W\left(\Gamma,\square\right)\left[R\right].
\end{equation}
Here $W\left(\Gamma,\square\right)\left[R\right]$ is the value of
$W\left(\Gamma,\square\right)$ evaluated at the representation $R$,
\begin{equation}
W\left(\Gamma,\square\right)\left[R\right]=\sum_{R_{+}=R\otimes\square}\frac{\dim R_{+}}{\dim R}\exp\left(-\frac{\varepsilon\Delta\mathcal{A}}{2}\left(C_{2}(R_{+})-C_{2}(R)\right)\right),\label{eq:WR}
\end{equation}
where $\Delta\mathcal{A}=\mathcal{A}(\Sigma_{2})-\mathcal{A}(\Sigma_{1})$.

First we consider the case when the gauge group is $SU(n)$. Suppose
that the Young diagram associated to the representation $R$ is $\left(\lambda_{1}\geq\lambda_{2}\geq\cdots\geq\lambda_{n}\geq0\right)$,
and becomes the Young diagram associated to the representation $R_{+}$
by adding a box in the $l$th row. From (\ref{eq:dimR}) and (\ref{eq:C2R}),
we obtain that 
\begin{eqnarray}
\frac{\dim R_{+}}{\dim R} & = & \prod_{i\neq r}\frac{\lambda_{i}-\left(\lambda_{l}+1\right)+l-i}{\lambda_{i}-\lambda_{l}+l-i},\label{eq:dimR+}\\
C_{2}(R_{+})-C_{2}(R) & = & 2\left(\lambda_{l}-l+1\right)+\frac{n^{2}-1-2|\lambda|}{n}.
\end{eqnarray}

It is interesting to notice that
\begin{equation}
\mathrm{Res}_{x=a_{1}+\hbar\left(\lambda_{l}^{(1)}+1-l\right)}\left(\frac{x+n\hbar}{\mathscr{Y}(x)[\Lambda]}\right)=\frac{\dim R_{+}}{\dim R}.\label{eq:Ydim}
\end{equation}
The appearance of the $\mathscr{Y}$-observable should not be surprising.
Recall that the physical meaning of the $\mathscr{Y}$-observable
is to add or remove a pointlike instanton. Hence, the four-dimensional
operator corresponding to $W\left(\Gamma,\square\right)[R]$ is
\begin{equation}
\frac{1}{2\pi\mathrm{i}}\oint dx\frac{x+n\hbar}{\mathscr{Y}(x)[\Lambda]}e^{-\varepsilon\Delta\mathcal{A}x}\exp\left(-\varepsilon\Delta\mathcal{A}\left(\frac{n^{2}-1}{2n}-\frac{1}{n}q\frac{\partial}{\partial q}\right)\right).
\end{equation}

For the case of $U(n)$, the equations (\ref{eq:dimR+}) and (\ref{eq:Ydim})
still hold. The difference between the Casimirs now is simpler 
\begin{equation}
C_{2}(\mathcal{R}_{+})-C_{2}(\mathcal{R})=2\left(\lambda_{l}-l+1\right)+n+2r.
\end{equation}
Hence, the four-dimensional operator corresponding to $W\left(\Gamma,\square\right)[R]$
is now
\begin{equation}
\frac{1}{2\pi\mathrm{i}}\oint dx\frac{x+n\hbar}{\mathscr{Y}(x)[\Lambda]}\exp\left(-\varepsilon\Delta\mathcal{A}\left(x+\frac{n}{2}\right)\right).
\end{equation}

\section{Discussions\label{sec:Dis}}

In this paper, we study a generalization of the correspondence between
four-dimensional $\mathcal{N}=2$ supersymmetric $U(N)$ gauge theory
with $2N-2$ fundamental hypermultiplets and A-type topological string
theory on $S^{2}$. In our correspondence, the partition function
of the four-dimensional $U(N)$ gauge theory with $2N$ fundamental
hypermultiplets at a suitable nongeneric point of the parameter space
is related to the partition function of two-dimensional Yang-Mills
theory on $S^{2}$. We also study the expectation value of a Wilson
loop operator in the fundamental representation in the two-dimensional
Yang-Mills theory. The corresponding operator in the four-dimensional
theory can be found for the fundamental representation. It appears
that the correspondence is more complicated than the old correspondence
in \cite{Losev:2003py,Marshakov:2006ii,Nekrasov:2009zza}.

The relation between four-dimensional supersymmetric gauge theory
and two-dimensional Yang-Mills theory on $S^{2}$ was discovered in
many other places. For example, the supersymmetric Wilson loops restricted
to an $S^{2}$ submanifold of four-dimensional space in $\mathcal{N}=4$
supersymmetric Yang-Mills theory \cite{Drukker:2007yx,Giombi:2009ds}
can be consistently truncated to a two-dimensional Yang-Mills theory
on $S^{2}$. However, the number of supersymmetry in four-dimensional
gauge theory and the way to identify the Wilson loop operator in their
work is quite different from our story. One other similar relation
is the identification of the superconformal index of a class of four-dimensional
$\mathcal{N}=2$ theories with a deformation of two-dimensional Yang-Mills
theory on punctured Riemann surfaces \cite{Gadde:2011ik}. However,
in their correspondence, the four-dimensional gauge theory is a complicated
quiver theory, and there are necessarily a number of punctures in
the Riemann surface. Hence, all these old relations are indeed different
from ours.

So far, the correspondence discussed in this paper is only a mathematical
coincidence of two different partition functions. It will be nice
if one can embed our correspondence into a string theory setup and
provide a physical interpretation of the results we have got. The
procedure (\ref{eq:condition2}) and (\ref{eq:condition1}) is similar
to the approach to introduce surface operators or vortices in the
previous discussions of AGT correspondence, and one may effectively
describe the surface operator as some two-dimensional gauge theory.
One may wonder whether the two-dimensional Yang-Mills theory we discuss
is somehow related to the gauge theory in this construction. However,
we would like to point out that this is not the case. Notice that
if we want to have a surface operator in a $U(N)$ gauge theory, we
can consider a two-dimensional gauge theory coupled to the $U(N)$
gauge theory, or we can start with a $U(N)\times U(N^{\prime})$ theory
and tune the Coulomb moduli in the $U(N^{\prime})$ part of the theory.
Furthermore, in this case, the two-dimensional gauge theory lives
inside the spacetime of the four-dimensional gauge theory. Instead,
we suggest that the proper physical origin of our result should come
from the compactification of little string theory. The four-dimensional
gauge theory and the two-dimensional Yang-Mills theory live in the
perpendicular spaces. This is also the case for the old correspondence
between supersymmetric gauge theory and topological string theory
\cite{Losev:2003py,Marshakov:2006ii}. 

There are many open problems which remain to be answered.

First, we only studied the Wilson loop operator which is inserted
in the two-dimensional Yang-Mills theory in the fundamental representation.
We can insert Wilson loop operators in arbitrary representations of
the gauge group and define a quantity similar to (\ref{eq:WR}),
\begin{equation}
W\left(\Gamma,R_{\Gamma}\right)\left[R\right]=\sum_{R_{+}}\frac{\dim R_{+}}{\dim R}\exp\left(-\frac{\varepsilon\Delta\mathcal{A}}{2}\left(C_{2}(R_{+})-C_{2}(R)\right)\right)\mathfrak{N}\left(R\otimes R_{\Gamma},R_{+}\right).
\end{equation}
Now $\mathfrak{N}\left(R\otimes R_{\Gamma},R_{+}\right)$ is more
complicated. What are the corresponding four-dimensional operators?

Second, we only consider the first nontrivial simplification of the
instanton partition function at a nongeneric point of the parameter
space in this paper. It is natural to extend our analysis to the cases
\begin{equation}
a_{1}=m_{1}+n_{1}\hbar=m_{2}+n_{1}\hbar,\quad a_{2}=m_{3}+n_{2}\hbar=m_{4}+n_{2}\hbar,\quad a_{3}=m_{5}=m_{6},\cdots,a_{N}=m_{2N-1}=m_{2N}.
\end{equation}
Then the length of the partition $\lambda^{(1)}$ is at most $n_{1}$,
the length of the partition $\lambda^{(2)}$ is at most $n_{2}$,
while all the other partitions are empty. Similar to the case discussed
in the paper, there are many cancellations in the measure. The resulting
measure is
\begin{eqnarray}
\mu & = & \left(\prod_{1\leq i<j\leq n_{1}}\frac{\lambda_{i}^{(1)}-\lambda_{j}^{(1)}+j-i}{j-i}\right)^{2}\left(\prod_{1\leq i<j\leq n_{2}}\frac{\lambda_{i}^{(2)}-\lambda_{j}^{(2)}+j-i}{j-i}\right)^{2}\times\nonumber \\
 & \times & \left(\prod_{i=1}^{n_{1}}\prod_{j=1}^{n_{2}}\frac{a_{1}-a_{2}+\hbar\left(\lambda_{i}^{(1)}-\lambda_{j}^{(2)}+j-i\right)}{a_{1}-a_{2}+\hbar\left(j-i\right)}\right)^{2}\times\nonumber \\
 & \times & \left(\prod_{i=1}^{n_{1}}\frac{\Gamma\left(\frac{a_{1}-a_{2}}{\hbar}+n_{2}+1+\lambda_{i}^{(1)}-i\right)}{\Gamma\left(\frac{a_{1}-a_{2}}{\hbar}+n_{2}+1-i\right)}\right)^{2}\left(\prod_{i=1}^{n_{2}}\frac{\Gamma\left(\frac{a_{2}-a_{1}}{\hbar}+n_{1}+1+\lambda_{i}^{(2)}-i\right)}{\Gamma\left(\frac{a_{2}-a_{1}}{\hbar}+n_{1}+1-i\right)}\right)^{2}.
\end{eqnarray}
What is the physical interpretation of this partition function?

\begin{acknowledgments}
Research was supported in part by the National Science Foundation Grant No. PHY 1404446. X.Z. has greatly benefited from discussions with Nikita Nekrasov. X.Z. would also like to thank Alex DiRe, Saebyeok Jeong, Naveen Prabhakar, Dan Xie, Wenbin Yan, and Peng Zhao for discussions.
\end{acknowledgments}

\bibliographystyle{apsrev4-1}
\bibliography{SW-YM2}

\begin{thebibliography}{17}%
\makeatletter
\providecommand \@ifxundefined [1]{%
 \@ifx{#1\undefined}
}%
\providecommand \@ifnum [1]{%
 \ifnum #1\expandafter \@firstoftwo
 \else \expandafter \@secondoftwo
 \fi
}%
\providecommand \@ifx [1]{%
 \ifx #1\expandafter \@firstoftwo
 \else \expandafter \@secondoftwo
 \fi
}%
\providecommand \natexlab [1]{#1}%
\providecommand \enquote  [1]{``#1''}%
\providecommand \bibnamefont  [1]{#1}%
\providecommand \bibfnamefont [1]{#1}%
\providecommand \citenamefont [1]{#1}%
\providecommand \href@noop [0]{\@secondoftwo}%
\providecommand \href [0]{\begingroup \@sanitize@url \@href}%
\providecommand \@href[1]{\@@startlink{#1}\@@href}%
\providecommand \@@href[1]{\endgroup#1\@@endlink}%
\providecommand \@sanitize@url [0]{\catcode `\\12\catcode `\$12\catcode
  `\&12\catcode `\#12\catcode `\^12\catcode `\_12\catcode `\%12\relax}%
\providecommand \@@startlink[1]{}%
\providecommand \@@endlink[0]{}%
\providecommand \url  [0]{\begingroup\@sanitize@url \@url }%
\providecommand \@url [1]{\endgroup\@href {#1}{\urlprefix }}%
\providecommand \urlprefix  [0]{URL }%
\providecommand \Eprint [0]{\href }%
\providecommand \doibase [0]{http://dx.doi.org/}%
\providecommand \selectlanguage [0]{\@gobble}%
\providecommand \bibinfo  [0]{\@secondoftwo}%
\providecommand \bibfield  [0]{\@secondoftwo}%
\providecommand \translation [1]{[#1]}%
\providecommand \BibitemOpen [0]{}%
\providecommand \bibitemStop [0]{}%
\providecommand \bibitemNoStop [0]{.\EOS\space}%
\providecommand \EOS [0]{\spacefactor3000\relax}%
\providecommand \BibitemShut  [1]{\csname bibitem#1\endcsname}%
\let\auto@bib@innerbib\@empty
\bibitem [{\citenamefont {Seiberg}\ and\ \citenamefont
  {Witten}(1994{\natexlab{a}})}]{Seiberg:1994rs}%
  \BibitemOpen
  \bibfield  {author} {\bibinfo {author} {\bibfnamefont {N.}~\bibnamefont
  {Seiberg}}\ and\ \bibinfo {author} {\bibfnamefont {E.}~\bibnamefont
  {Witten}},\ }\href {\doibase 10.1016/0550-3213(94)90124-4} {\bibfield
  {journal} {\bibinfo  {journal} {Nucl.Phys.}\ }\textbf {\bibinfo {volume}
  {B426}},\ \bibinfo {pages} {19} (\bibinfo {year} {1994}{\natexlab{a}})},\
  \Eprint {http://arxiv.org/abs/hep-th/9407087} {arXiv:hep-th/9407087 [hep-th]}
  \BibitemShut {NoStop}%
\bibitem [{\citenamefont {Seiberg}\ and\ \citenamefont
  {Witten}(1994{\natexlab{b}})}]{Seiberg:1994aj}%
  \BibitemOpen
  \bibfield  {author} {\bibinfo {author} {\bibfnamefont {N.}~\bibnamefont
  {Seiberg}}\ and\ \bibinfo {author} {\bibfnamefont {E.}~\bibnamefont
  {Witten}},\ }\href {\doibase 10.1016/0550-3213(94)90214-3} {\bibfield
  {journal} {\bibinfo  {journal} {Nucl.Phys.}\ }\textbf {\bibinfo {volume}
  {B431}},\ \bibinfo {pages} {484} (\bibinfo {year} {1994}{\natexlab{b}})},\
  \Eprint {http://arxiv.org/abs/hep-th/9408099} {arXiv:hep-th/9408099 [hep-th]}
  \BibitemShut {NoStop}%
\bibitem [{\citenamefont {Tachikawa}(2013)}]{Tachikawa:2013kta}%
  \BibitemOpen
  \bibfield  {author} {\bibinfo {author} {\bibfnamefont {Y.}~\bibnamefont
  {Tachikawa}},\ }in\ \href {\doibase 10.1007/978-3-319-08822-8} {\emph
  {\bibinfo {booktitle} {{Lecture Notes in Physics, vol. 890, 2014}}}},\ Vol.\
  \bibinfo {volume} {890}\ (\bibinfo {year} {2013})\ p.\ \bibinfo {pages}
  {2014},\ \Eprint {http://arxiv.org/abs/1312.2684} {arXiv:1312.2684 [hep-th]}
  \BibitemShut {NoStop}%
\bibitem [{\citenamefont {Nekrasov}(2004)}]{Nekrasov:2002qd}%
  \BibitemOpen
  \bibfield  {author} {\bibinfo {author} {\bibfnamefont {N.~A.}\ \bibnamefont
  {Nekrasov}},\ }\href {\doibase 10.4310/ATMP.2003.v7.n5.a4} {\bibfield
  {journal} {\bibinfo  {journal} {Adv.Theor.Math.Phys.}\ }\textbf {\bibinfo
  {volume} {7}},\ \bibinfo {pages} {831} (\bibinfo {year} {2004})},\ \Eprint
  {http://arxiv.org/abs/hep-th/0206161} {arXiv:hep-th/0206161 [hep-th]}
  \BibitemShut {NoStop}%
\bibitem [{\citenamefont {Festuccia}\ and\ \citenamefont
  {Seiberg}(2011)}]{Festuccia:2011ws}%
  \BibitemOpen
  \bibfield  {author} {\bibinfo {author} {\bibfnamefont {G.}~\bibnamefont
  {Festuccia}}\ and\ \bibinfo {author} {\bibfnamefont {N.}~\bibnamefont
  {Seiberg}},\ }\href {\doibase 10.1007/JHEP06(2011)114} {\bibfield  {journal}
  {\bibinfo  {journal} {JHEP}\ }\textbf {\bibinfo {volume} {06}},\ \bibinfo
  {pages} {114} (\bibinfo {year} {2011})},\ \Eprint
  {http://arxiv.org/abs/1105.0689} {arXiv:1105.0689 [hep-th]} \BibitemShut
  {NoStop}%
\bibitem [{\citenamefont {Losev}\ \emph {et~al.}(2003)\citenamefont {Losev},
  \citenamefont {Marshakov},\ and\ \citenamefont {Nekrasov}}]{Losev:2003py}%
  \BibitemOpen
  \bibfield  {author} {\bibinfo {author} {\bibfnamefont {A.~S.}\ \bibnamefont
  {Losev}}, \bibinfo {author} {\bibfnamefont {A.}~\bibnamefont {Marshakov}}, \
  and\ \bibinfo {author} {\bibfnamefont {N.~A.}\ \bibnamefont {Nekrasov}},\
  }\href@noop {} {\  (\bibinfo {year} {2003})},\ \Eprint
  {http://arxiv.org/abs/hep-th/0302191} {arXiv:hep-th/0302191 [hep-th]}
  \BibitemShut {NoStop}%
\bibitem [{\citenamefont {Marshakov}\ and\ \citenamefont
  {Nekrasov}(2007)}]{Marshakov:2006ii}%
  \BibitemOpen
  \bibfield  {author} {\bibinfo {author} {\bibfnamefont {A.}~\bibnamefont
  {Marshakov}}\ and\ \bibinfo {author} {\bibfnamefont {N.}~\bibnamefont
  {Nekrasov}},\ }\href {\doibase 10.1088/1126-6708/2007/01/104} {\bibfield
  {journal} {\bibinfo  {journal} {JHEP}\ }\textbf {\bibinfo {volume} {0701}},\
  \bibinfo {pages} {104} (\bibinfo {year} {2007})},\ \Eprint
  {http://arxiv.org/abs/hep-th/0612019} {arXiv:hep-th/0612019 [hep-th]}
  \BibitemShut {NoStop}%
\bibitem [{\citenamefont {Nekrasov}(2009)}]{Nekrasov:2009zza}%
  \BibitemOpen
  \bibfield  {author} {\bibinfo {author} {\bibfnamefont {N.~A.}\ \bibnamefont
  {Nekrasov}},\ }\href {\doibase 10.1007/s11005-009-0312-9} {\bibfield
  {journal} {\bibinfo  {journal} {Lett.Math.Phys.}\ }\textbf {\bibinfo {volume}
  {88}},\ \bibinfo {pages} {207} (\bibinfo {year} {2009})}\BibitemShut
  {NoStop}%
\bibitem [{\citenamefont {Migdal}(1975)}]{Migdal:1975zg}%
  \BibitemOpen
  \bibfield  {author} {\bibinfo {author} {\bibfnamefont {A.~A.}\ \bibnamefont
  {Migdal}},\ }\href@noop {} {\bibfield  {journal} {\bibinfo  {journal} {Sov.
  Phys. JETP}\ }\textbf {\bibinfo {volume} {42}},\ \bibinfo {pages} {413}
  (\bibinfo {year} {1975})},\ \bibinfo {note} {[Zh. Eksp. Teor.
  Fiz.69,810(1975)]}\BibitemShut {NoStop}%
\bibitem [{\citenamefont {Cordes}\ \emph {et~al.}(1995)\citenamefont {Cordes},
  \citenamefont {Moore},\ and\ \citenamefont {Ramgoolam}}]{Cordes:1994fc}%
  \BibitemOpen
  \bibfield  {author} {\bibinfo {author} {\bibfnamefont {S.}~\bibnamefont
  {Cordes}}, \bibinfo {author} {\bibfnamefont {G.~W.}\ \bibnamefont {Moore}}, \
  and\ \bibinfo {author} {\bibfnamefont {S.}~\bibnamefont {Ramgoolam}},\ }\href
  {\doibase 10.1016/0920-5632(95)00434-B} {\bibfield  {journal} {\bibinfo
  {journal} {Nucl.Phys.Proc.Suppl.}\ }\textbf {\bibinfo {volume} {41}},\
  \bibinfo {pages} {184} (\bibinfo {year} {1995})},\ \Eprint
  {http://arxiv.org/abs/hep-th/9411210} {arXiv:hep-th/9411210 [hep-th]}
  \BibitemShut {NoStop}%
\bibitem [{\citenamefont {Nekrasov}\ and\ \citenamefont
  {Pestun}(2012)}]{Nekrasov:2012xe}%
  \BibitemOpen
  \bibfield  {author} {\bibinfo {author} {\bibfnamefont {N.}~\bibnamefont
  {Nekrasov}}\ and\ \bibinfo {author} {\bibfnamefont {V.}~\bibnamefont
  {Pestun}},\ }\href@noop {} {\  (\bibinfo {year} {2012})},\ \Eprint
  {http://arxiv.org/abs/1211.2240} {arXiv:1211.2240 [hep-th]} \BibitemShut
  {NoStop}%
\bibitem [{\citenamefont {Nekrasov}(2016)}]{Nekrasov:2015wsu}%
  \BibitemOpen
  \bibfield  {author} {\bibinfo {author} {\bibfnamefont {N.}~\bibnamefont
  {Nekrasov}},\ }\href {\doibase 10.1007/JHEP03(2016)181} {\bibfield  {journal}
  {\bibinfo  {journal} {JHEP}\ }\textbf {\bibinfo {volume} {03}},\ \bibinfo
  {pages} {181} (\bibinfo {year} {2016})},\ \Eprint
  {http://arxiv.org/abs/1512.05388} {arXiv:1512.05388 [hep-th]} \BibitemShut
  {NoStop}%
\bibitem [{\citenamefont {Rusakov}(1990)}]{Rusakov:1990rs}%
  \BibitemOpen
  \bibfield  {author} {\bibinfo {author} {\bibfnamefont {B.~E.}\ \bibnamefont
  {Rusakov}},\ }\href {\doibase 10.1142/S0217732390000780} {\bibfield
  {journal} {\bibinfo  {journal} {Mod. Phys. Lett.}\ }\textbf {\bibinfo
  {volume} {A5}},\ \bibinfo {pages} {693} (\bibinfo {year} {1990})}\BibitemShut
  {NoStop}%
\bibitem [{\citenamefont {Witten}(1991)}]{Witten:1991we}%
  \BibitemOpen
  \bibfield  {author} {\bibinfo {author} {\bibfnamefont {E.}~\bibnamefont
  {Witten}},\ }\href {\doibase 10.1007/BF02100009} {\bibfield  {journal}
  {\bibinfo  {journal} {Commun. Math. Phys.}\ }\textbf {\bibinfo {volume}
  {141}},\ \bibinfo {pages} {153} (\bibinfo {year} {1991})}\BibitemShut
  {NoStop}%
\bibitem [{\citenamefont {Drukker}\ \emph {et~al.}(2008)\citenamefont
  {Drukker}, \citenamefont {Giombi}, \citenamefont {Ricci},\ and\ \citenamefont
  {Trancanelli}}]{Drukker:2007yx}%
  \BibitemOpen
  \bibfield  {author} {\bibinfo {author} {\bibfnamefont {N.}~\bibnamefont
  {Drukker}}, \bibinfo {author} {\bibfnamefont {S.}~\bibnamefont {Giombi}},
  \bibinfo {author} {\bibfnamefont {R.}~\bibnamefont {Ricci}}, \ and\ \bibinfo
  {author} {\bibfnamefont {D.}~\bibnamefont {Trancanelli}},\ }\href {\doibase
  10.1103/PhysRevD.77.047901} {\bibfield  {journal} {\bibinfo  {journal} {Phys.
  Rev.}\ }\textbf {\bibinfo {volume} {D77}},\ \bibinfo {pages} {047901}
  (\bibinfo {year} {2008})},\ \Eprint {http://arxiv.org/abs/0707.2699}
  {arXiv:0707.2699 [hep-th]} \BibitemShut {NoStop}%
\bibitem [{\citenamefont {Giombi}\ and\ \citenamefont
  {Pestun}(2010)}]{Giombi:2009ds}%
  \BibitemOpen
  \bibfield  {author} {\bibinfo {author} {\bibfnamefont {S.}~\bibnamefont
  {Giombi}}\ and\ \bibinfo {author} {\bibfnamefont {V.}~\bibnamefont
  {Pestun}},\ }\href {\doibase 10.1007/JHEP10(2010)033} {\bibfield  {journal}
  {\bibinfo  {journal} {JHEP}\ }\textbf {\bibinfo {volume} {10}},\ \bibinfo
  {pages} {033} (\bibinfo {year} {2010})},\ \Eprint
  {http://arxiv.org/abs/0906.1572} {arXiv:0906.1572 [hep-th]} \BibitemShut
  {NoStop}%
\bibitem [{\citenamefont {Gadde}\ \emph {et~al.}(2011)\citenamefont {Gadde},
  \citenamefont {Rastelli}, \citenamefont {Razamat},\ and\ \citenamefont
  {Yan}}]{Gadde:2011ik}%
  \BibitemOpen
  \bibfield  {author} {\bibinfo {author} {\bibfnamefont {A.}~\bibnamefont
  {Gadde}}, \bibinfo {author} {\bibfnamefont {L.}~\bibnamefont {Rastelli}},
  \bibinfo {author} {\bibfnamefont {S.~S.}\ \bibnamefont {Razamat}}, \ and\
  \bibinfo {author} {\bibfnamefont {W.}~\bibnamefont {Yan}},\ }\href {\doibase
  10.1103/PhysRevLett.106.241602} {\bibfield  {journal} {\bibinfo  {journal}
  {Phys. Rev. Lett.}\ }\textbf {\bibinfo {volume} {106}},\ \bibinfo {pages}
  {241602} (\bibinfo {year} {2011})},\ \Eprint {http://arxiv.org/abs/1104.3850}
  {arXiv:1104.3850 [hep-th]} \BibitemShut {NoStop}%
\end{thebibliography}%

\end{document}